\documentclass[aps,prl,twocolumn,superscriptaddress]{revtex4-1}
\usepackage{graphicx}
\usepackage{bbold}
\usepackage{amsmath}
\usepackage{upgreek}
\usepackage{xcolor}
\usepackage[utf8]{inputenc}
\usepackage{graphicx}
\usepackage[margin=1in]{geometry}
\usepackage{amsmath}
\usepackage{amssymb}

\newcommand{\ket}[1]{{\left| {#1} \right\rangle}}

\newcommand{\ketbra}[2]{{\left| {#1} \right\rangle
    \!\!\left\langle{#2} \right|}}

\newcommand{\mrm}[1]{\mathrm{#1}}

\newcommand{\Vket}[1]{{\left| {#1} \right\rangle}_\mathrm{V}}
\newcommand{\Hket}[1]{{\left| {#1} \right\rangle}_\mathrm{H}}

\newcommand{\atomicPauli}[3]{{\hat{s}_{\mathtt{\underline{#2}\hspace{0.2mm}\underline{#3}}}^{(#1)}}}
\newcommand{\atomicToSpin}[2]{{\hat{T}_{\mathtt{\underline{#1}\hspace{0.1mm}\underline{#2}}}}}
\newcommand{\atomicRabi}[2]{{\Omega_{\mathtt{\underline{#1}\hspace{0.1mm}\underline{#2}}}}}

\usepackage{bbm}
\newcommand{\idty}{\mathbbm{1}}
\newcommand{\XX}{\textsc{xx}}
\newcommand{\ZZ}{\textsc{zz}}

\newcommand{\Aket}[1]{\ket{\mathtt{\underline{#1}}}}

\begin{document}

\title{Polyqubit quantum processing}
\author{Wesley C. Campbell}
\affiliation{Department of Physics and Astronomy, Los Angeles, California 90095, USA}
\affiliation{UCLA Center for Quantum Science and Engineering, University of California – Los Angeles, Los Angeles, California 90095, USA}
\affiliation{Challenge Institute for Quantum Computation, University of California – Los Angeles, Los Angeles, California 90095, USA}
\author{Eric R. Hudson}
\affiliation{Department of Physics and Astronomy, Los Angeles, California 90095, USA}
\affiliation{UCLA Center for Quantum Science and Engineering, University of California – Los Angeles, Los Angeles, California 90095, USA}
\affiliation{Challenge Institute for Quantum Computation, University of California – Los Angeles, Los Angeles, California 90095, USA}

\begin{abstract}
We describe the encoding of multiple qubits per atom in trapped atom quantum processors and methods for performing both intra- and inter-atomic gates on participant qubits without disturbing the spectator qubits stored in the same atoms.
We also introduce techniques for selective state preparation and measurement of individual qubits that leave the information encoded in the other qubits intact, a capability required for qubit quantum error correction. 
The additional internal states needed for polyqubit processing are already present in atomic processors, suggesting that the resource cost associated with this multiplicative increase in qubit number could be a good bargain in the short to medium term.
\end{abstract}

\date{\today}
\maketitle

Quantum technologies use quantum objects, such as atoms, photons, phonons, and electrons to house, transport, and process quantum information.
While some applications, like certified randomness~\cite{Bruner2014}, explicitly utilize the nonlocality of quantum entanglement as a resource, most quantum computing algorithms do not.
For these applications, it is therefore not necessary that each qubit be encoded in a physically distinct object.  
Given that the atoms currently used to host qubits have many accessible internal states, it is natural to ask: can the computational power of atomic processors be increased, at an acceptable resource cost, by defining multiple qubits within each atom?

As an example, a problem requiring a Hilbert space of dimension $2^n$ could in principle be processed by a single atom, a \emph{unary encoding}, with a large-enough number of internal states.
Though this has the advantage of requiring only one atom, both the information storage and system control resources grow exponentially with problem size, and it has been shown that scalable quantum computing is not possible with a unary processor~\cite{Ekert1998Quantum,Blume-Kohout2002Climbing}. 

At the other extreme is the current pagadigm, in which each atom hosts a single qubit, a \emph{monoqubit encoding}. 
Here, the computational Hilbert space can be factored as a tensor product of $n$, 2-dimensional qubit subspaces as $\mathcal{H}_\mrm{comp}^{(\mathrm{qubit})} = \bigotimes_{i = 1}^n \mathcal{H}_2^{(i)}$.
Since the control resources, which essentially dictate this decomposition~\cite{Zanardi2001Virtual}, do not grow exponentially with problem size, such an encoding can potentially be used for scalable quantum processing.
However, at present, a number of technical considerations, including practical bounds on the number of shared Bosonic modes, laser paths, trap zones, occupied tweezer sites, frequency modulators, and/or other controls
~\cite{Bruzewicz2019TrappedIon,Kaufman2021Quantum,Monroe2013Scaling},
limit the number of \emph{atoms} that can be reliably employed in realized processors to tens to hundreds. 
Therefore, in the present qubit-host-limited (QHL) era it may be beneficial to encode a small
number ($p>1$) of qubits per atom, provided the control resources can be managed.
\begin{figure}
    \centering
    \includegraphics[width = 0.75\columnwidth]{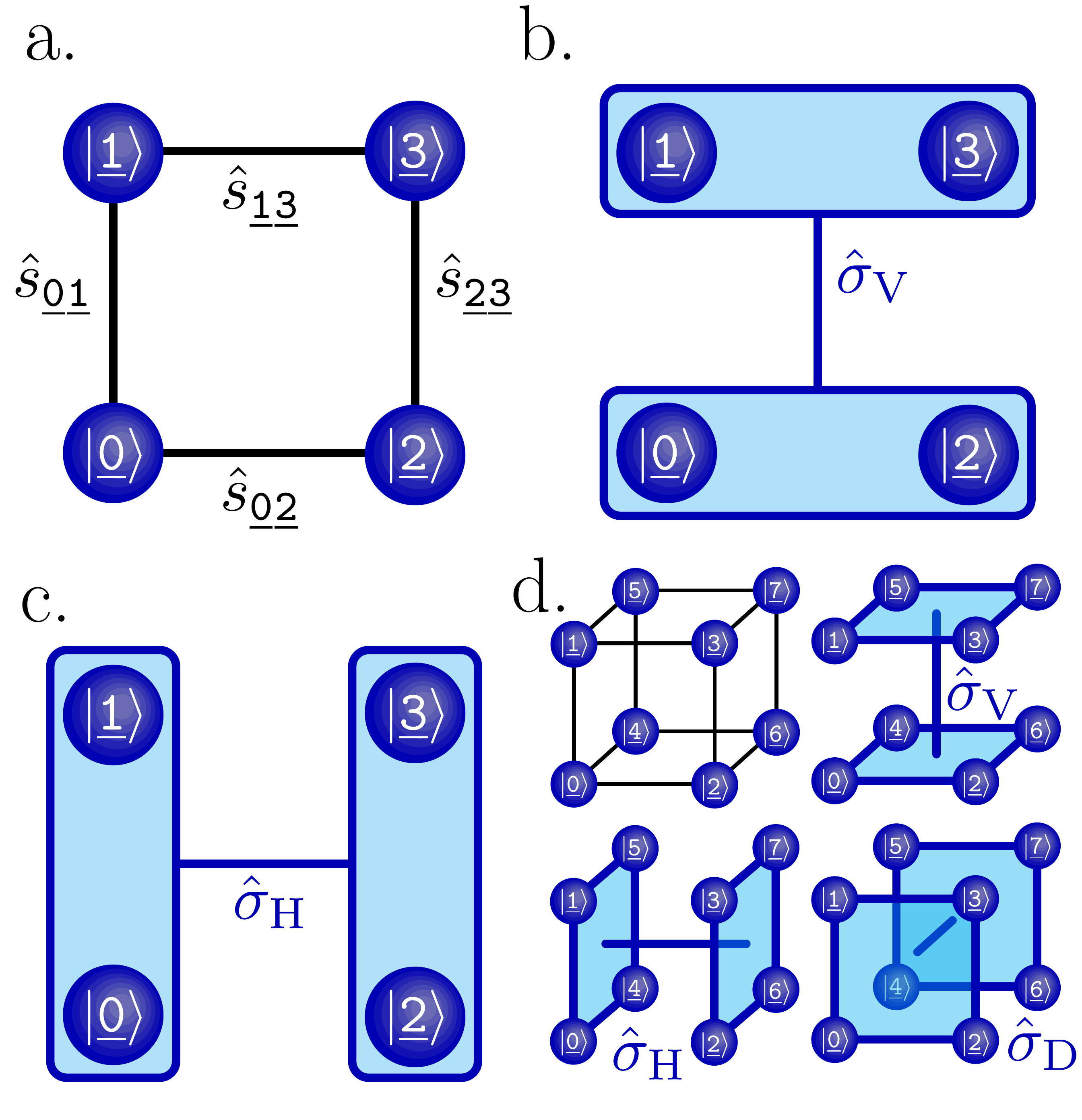}
    \caption{(a-c) A $p=2$ polyqubit encoded in four atomic eigenstates.  (a) Each atomic Pauli operator $\atomicPauli{i}{m}{n}$ acts on only two atomic eigenstates.  (b) ``Vertical'' $(\hat{\sigma}_\mrm{V}^{(i)})$ and (c) ``Horizontal'' $(\hat{\sigma}_\mrm{H}^{(i)})$ qubit operators connect qubit states. (d) A $p=3$ polyqubit can be encoded in eight atomic eigenstates to add a ``Depth'' ($\mathrm{D}$) qubit, $\mathcal{H}_{2^3} = \mathcal{H}_2^{(\mathrm{D})} \otimes \mathcal{H}_2^{(\mathrm{H})} \otimes \mathcal{H}_2^{(\mathrm{V})}$.  Each qubit operator is built from $2^{p-1}$ atomic Pauli operators. }
    \label{fig:StatesBallAndStick}
\end{figure}

Here, we show how atomic processors may be built from a \emph{polyqubit encoding} with a Hilbert space composed as
$\mathcal{H}_\mrm{comp}^{(\mathrm{polyqubit})} =\bigotimes_{i = 1}^{n/p} \mathcal{H}_{2^p}^{(i)}$.
The main distinction between this polyqubit processing and non-binary (qudit) quantum processing is that polyqubit processing requires the ability to perform state preparation and measurement (SPAM) and gates on participant qubits without disturbing spectator qubits stored in the \emph{same} atom.
This operational access allows the atomic subspace,
$\mathcal{H}_{2^p}$, to be factored into a tensor product of qubit subspaces as $\mathcal{H}_{2^p}^{(i)} = \bigotimes_{j=1}^{p}\mathcal{H}_2^{(i,j)}$~\cite{Zanardi2001Virtual},
whereas an equally-dimensioned qudit does not in general admit this factorization.
This enables a polyqubit machine to use standard, binary quantum algorithms without modification, including qubit quantum error correction (QEC).

A $p$-polyqubit encoded in an atomic subspace, \textit{i.e.}\ $p$ qubits stored in one atom, can be visualized as a $p$-dimensional hypercube graph with atomic eigenstates on the vertices and atomic Pauli operations on each edge (see Fig.~\ref{fig:StatesBallAndStick}).  
From the number of edges, it is clear that a polyqubit encoding requires $p\,2^{p-1}$ sets of \emph{atomic} Pauli operators that act on only two atomic eigenstates.
Though this control parameter cost is exponential in $p$, with fixed $p$ the processor itself scales with problem size by increasing the number of polyqubits as $n/p$.  
Thus, by keeping $p$ small enough to manage control costs, polyqubit encoding provides a multiplicative boost to 
QHL processors.
In what follows, we describe how $p$ qubits can be encoded into $2^p$ states of each single atom in a trapped ion quantum processor and used with currently available technology.

As an example of $p=2$ polyqubit processing, we consider a linear chain of atomic ions whose motion in a particular normal mode serves as a bus \cite{Bruzewicz2019TrappedIon}.
The atomic ions are assumed to have four, long-lived internal states appropriate for quantum information storage, labeled with underscores for clarity $\{\Aket{0},\Aket{1},\Aket{2},\Aket{3}\}$ (see Fig.~\ref{fig:StatesBallAndStick}). 
These could be hyperfine or Zeeman levels of a ground or metastable electronic state or some combination thereof~\cite{Allcock2021}.  
We assume that the transitions between all pairs of states occur with unique frequencies, and that at least four of them can be driven to achieve quadrilateral connectivity, as shown in Fig.~\ref{fig:StatesBallAndStick}a. 
On this support, we define two qubits that we dub ``Horizontal'' ($\mathrm{H}$) and ``Vertical'' ($\mathrm{V}$), with the mapping:
\begin{align}
	&\Aket{0} \equiv  \Hket{0} \otimes \Vket{0}  \nonumber\\
	&\Aket{1} \equiv  \Hket{0} \otimes \Vket{1} \nonumber\\
	&\Aket{2} \equiv  \Hket{1} \otimes \Vket{0} \nonumber\\
	&\Aket{3} \equiv  \Hket{1} \otimes \Vket{1}
\end{align}
so that each polyqubit state $\Hket{x'}\otimes \Vket{x}$ is the atomic state index expressed in two-digit ($x' \, x$) binary.  
As this user-defined designation of states is arbitrary, results for $\mrm{H}$ and $\mrm{V}$ qubits are always interchangeable and we require no particular physical difference between the types.

SPAM of polyqubits proceeds as follows.
If all of the qubits in a polyencoded atom are to be initialized or read out, techniques from binary processing can be adopted with only slight modification.
For example, optical pumping with polarization- or frequency-controlled light can produce a single atomic state with high-purity~\cite{Christensen2021},
which can be subsequently manipulated by microwave or optical radiation to prepare any desired polyqubit state.
State detection of all the qubits in an atom can be accomplished by transferring the polyqubit to a metastable manifold and serially transferring each atomic state into the ground state, where laser-induced fluorescence (LIF) is used for detection~\cite{Low2020Practical}. 
Finding the atom in a single state, which is heralded by LIF, fully determines the value of all of the polyencoded qubits. 
This type of state detection is well known~\cite{two_qubit_encoding}.
\begin{figure}
    \centering
    \includegraphics[width = 0.8\columnwidth]{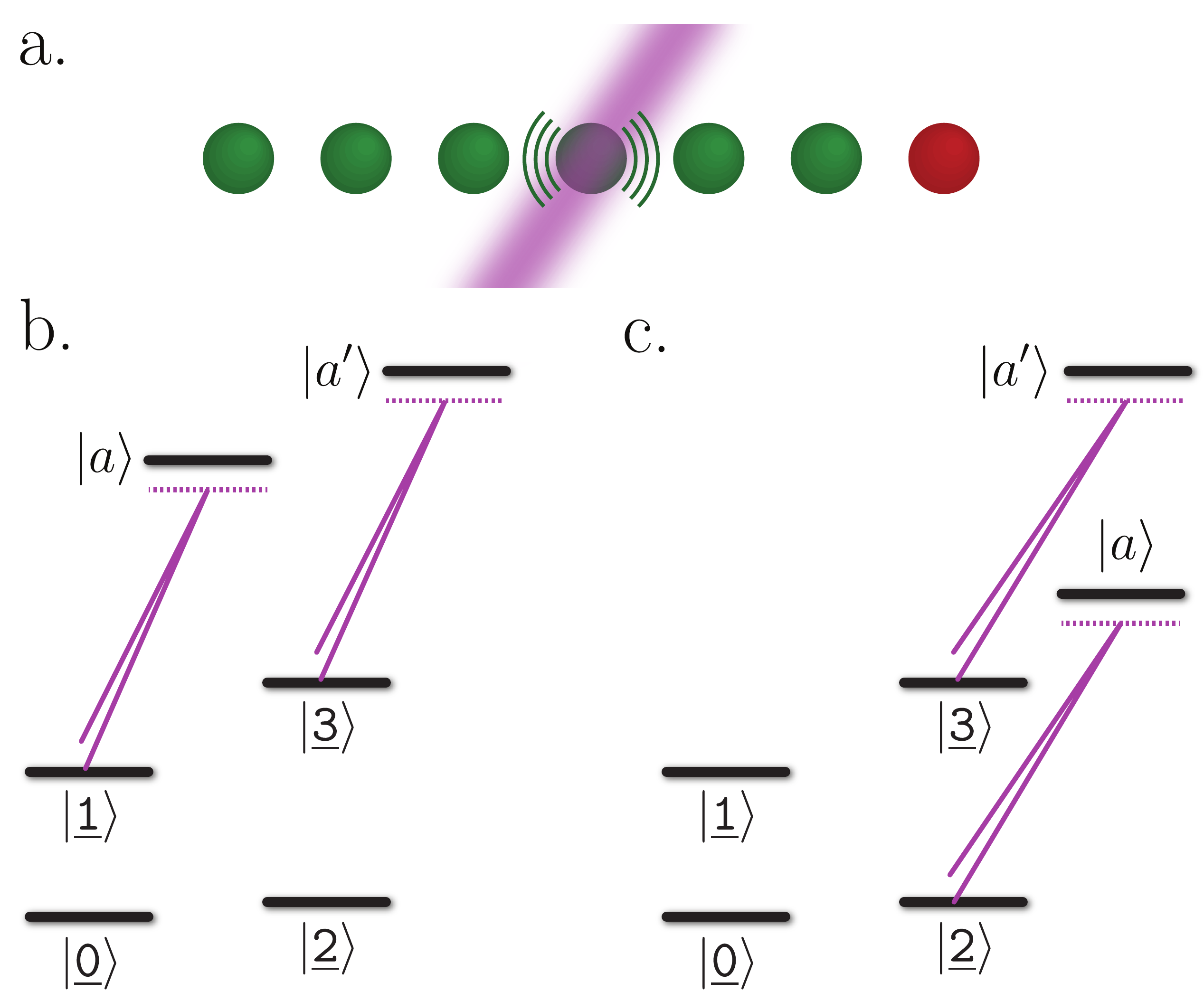}
    \caption{State detection of a qubit in a polyqubit-encoded ion (green) by laser shaking and subsequent motion readout using a co-trapped ancilla ion (red). An individually-addressed laser beam with a beatnotes at a normal mode frequency (purple) used for state detection of a (b) vertical or (c) horizontal qubit. This interaction can also be leveraged to perform inter-atomic \ZZ\ gates.}
    \label{fig:SPAM}
\end{figure}

For SPAM of an \emph{individual} qubit within a polyencoded atom, the participant qubit must be measured without disturbing the spectator qubits.
In general, this can be accomplished by a measurement that leaves the measured qubit in an eigenstate of the measurement (a quantum nondemolition measurement) and will typically require an ancilla ion.   

As an example, state detection of a single qubit in a $p=2$ polyencoded ion using a co-trapped ancilla ion could proceed as follows. 
First, a desired mode of motion of the trapped ion crystal is cooled near its ground state using, for example, the ancilla ion.
Next, a laser is used to add energy to this mode if and only if the participant qubit is in the particular state, as shown in Fig.~\ref{fig:SPAM}. 
As a qubit in a polyencoded atom corresponds to the population being localized in a \emph{manifold} of atomic states (\textit{e.g.}, $\Hket{1}$ indicates an ion whose state is in the manifold  $\{ \Aket{2}, \Aket{3} \}$ in Fig.~\ref{fig:StatesBallAndStick}), the laser interaction must be capable of adding motional energy without distinguishing between states within the manifold.
This can be achieved by \textit{e.g.}\ applying a motion-sensitive laser field that is modulated at the motional-mode frequency close to a narrow optical transition 
to produce an AC Stark shift of only the states in the participant qubit manifold~\cite{Baldwin2021HighFidelity}. 
By matching the strength and phase of this time-dependent Stark shift for each state in the manifold, energy is added without disturbing information encoded in the manifold. 
The ancilla ion can then be interrogated via standard techniques to determine if the motional energy of the ions has increased, thereby performing a projective measurement on only the participant qubit. 
Finally, state preparation of a participant qubit in a polyencoded atom can be accomplished by performing state detection followed by any necessary single-qubit rotation.

Alternatively, as described below, the complete gate set of the system presented here makes it possible to perform a two-qubit \textsc{swap} gate between the participant qubit and an initialized qubit in the ancilla ion.
Thus, the qubit can be written into the ancilla internal state for detection or the ancilla state written into the qubit for preparation.

The construction of single qubit gates for polyqubits can be intuited by expressing the qubit Pauli operators, $\hat{\sigma}_{\mrm{H}}^{(i)}$ and $\hat{\sigma}_{\mrm{V}}^{(i)}$ with $i \in \{X,Y,Z\}$,
in terms of \textit{atomic Pauli operators} -- \textit{i.e.}\ Pauli operators that act only on pairs of atomic states, denoted by $\atomicPauli{i}{m}{n}$ with dimension $2^p\times2^p$.
The atomic Pauli operators are constructed from the unitary $2 \times 2$ Pauli matrices $\hat{\sigma}^{(i)}$ (with $\ket{\uparrow}$ and $\ket{\downarrow}$ as the positive- and negative-eigenvalue eigenvectors of $\hat{\sigma}^{(Z)}$, respectively) using the
operator $\atomicToSpin{m}{n} \equiv \ketbra{\uparrow}{\mathtt{\underline{m}}} + \ketbra{\downarrow}{\mathtt{\underline{n}}}$ as $\atomicPauli{i}{m}{n} \equiv \atomicToSpin{m}{n}^\dagger \,\sigma^{(i)} \atomicToSpin{m}{n}$.
For the $p=2$ case shown in Fig.~\ref{fig:StatesBallAndStick}, we have
\begin{align}
	\hat{\sigma}_\mrm{H}^{(i)}  &= \atomicPauli{i}{0}{2} + \atomicPauli{i}{1}{3}\nonumber\\
	\hat{\sigma}_\mrm{V}^{(i)} &= \atomicPauli{i}{0}{1} + \atomicPauli{i}{2}{3}. \label{eq:AtomicPaulis}
\end{align}

The $\atomicPauli{i}{m}{n}$ are realized, just as they are for monoqubit encoding, by application of electromagnetic radiation near resonance, either in a direct or stimulated Raman fashion, with the $\Aket{m} \! \leftrightarrow \!\Aket{n}$ transition~\cite{Bruzewicz2019TrappedIon}.
Because $\left[ \atomicPauli{i}{m}{n}, \atomicPauli{i}{k}{l} \right] = 0$ if $\{\mathtt{\underline{m}}, \mathtt{\underline{n}} \} \cap \{\mathtt{\underline{k}}, \mathtt{\underline{l}} \} = \emptyset $, it is not strictly necessary to drive both transitions simultaneously for single qubit gates.  
It is, however, necessary that the pulse areas
of both transitions are matched and, in most cases, the relative phase controlled.
Therefore, it is likely most efficient to realize single-qubit gates via simultaneous, multi-tone modulation of a single optical source, and we will assume simultaneous implementation is used.  
Since many of the gate errors encountered in trapped ion processors (including optical phase, frequency, and amplitude errors) can be made common-mode to both in this case, such errors will appear on the particpant qubit without distrubing the spectator qubit, an assumption underlying most qubit QEC schemes.

There are two distinct types of two-qubit gates for polyqubits: intra-atomic and inter-atomic. 
Intra-atomic two-qubit gates only require single-atom operations~\cite{two_qubit_encoding}.
As an example, an intra-atomic \textsc{cnot} gate between the $\mrm{H}$ and $\mrm{V}$ qubits of a $p = 2$ polyqubit can be driven by simply applying a resonant $\pi$ pulse on the $\Aket{2}\leftrightarrow\Aket{3}$ transition along with a $\frac{\pi}{2}$ phase shift ($\hat{S}_\mathrm{H}$) on the $\mathrm{H}$ qubit, as 
\begin{align}
e^{i \frac{\pi}{4}}\exp \left(-i \frac{\pi}{4} \hat{\sigma}_\mathrm{H}^{(Z)} \right)\exp \left(-i \frac{\pi}{2}\atomicPauli{X}{2}{3}\right) = & \,\,\left( \begin{array}{cc} \idty & 0 \\ 0 & \sigma^{(X)}  \end{array} \right) \nonumber \\
= & \,\, \textsc{cnot}.
\end{align}
Similar ideas can be applied to produce inter-atomic multi-qubit gates involving $p>2$ qubits, including Deutsch and Toffoli gates \cite{Deutsch1989}, requiring only single-atom internal state transitions \cite{SI}.
Since these gates require only single atom operations, high fidelities and high speeds can be expected with current technology.  
\begin{figure}
    \centering
    \includegraphics[width = 0.4\textwidth]{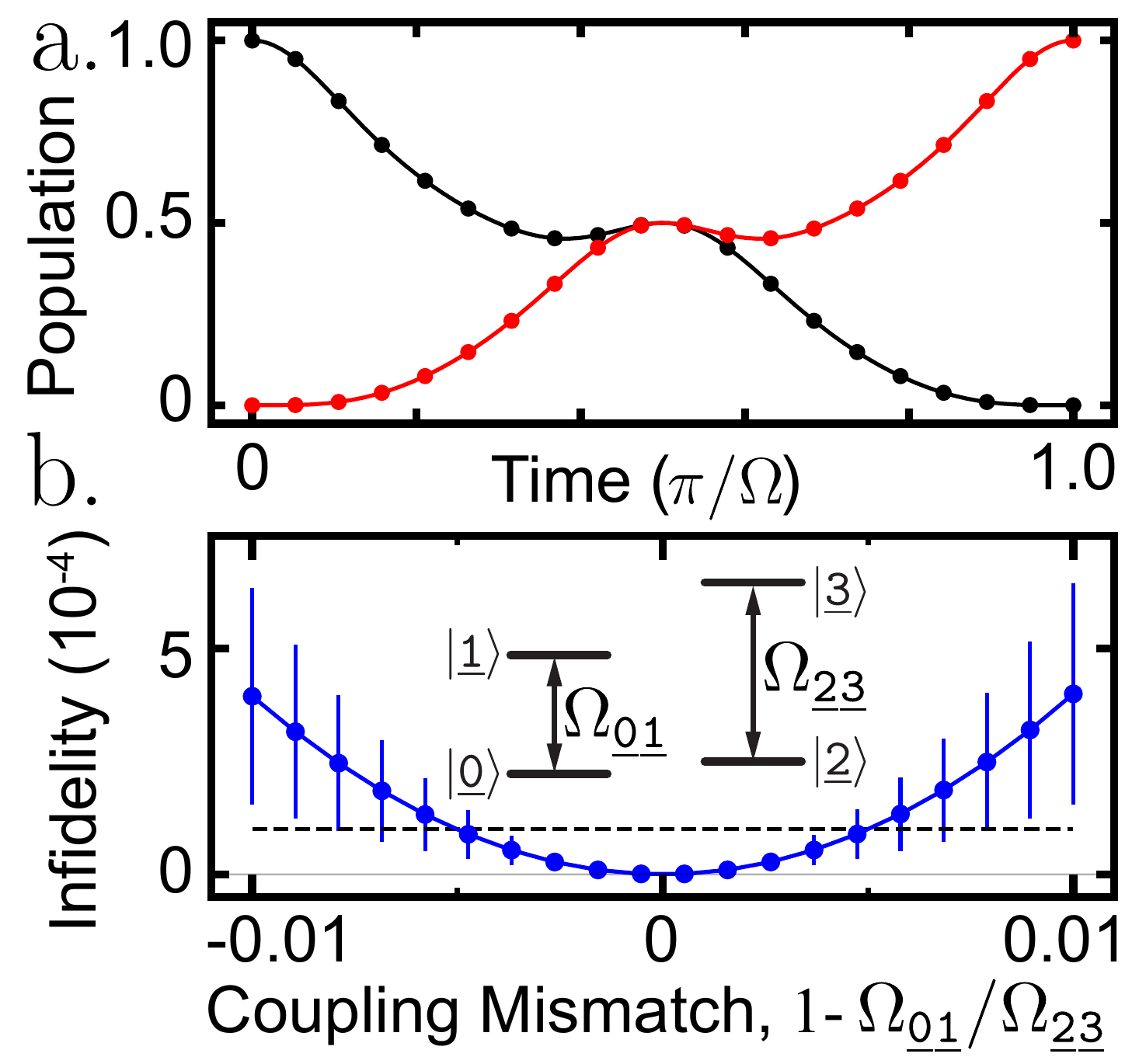}
    \caption{(a) Numerical calculation of the evolution during an inter-atomic M\o{}lmer-S\o{}rensen interaction between two polyencoded ions. 
    The dots are the population in the $\ket{0}_{d_1}\otimes\ket{0}_{d_2}$ (black) and $\ket{1}_{d_1}\otimes\ket{1}_{d_2}$ (red) participant qubit states.
    Each point is calculated with a randomly selected initial state for the spectator qubit.
    Solid lines show evolution of monoencoded ions under the same conditions.
    (b) Fidelity of the Bell state prepared at $\Omega t/\pi= 0.5$ as the Rabi frequencies of the two, motion-coupling atomic transitions are mismatched. One example encoding corresponding to $d_1=d_2=\mathrm{V}$ qubits is shown in the inset; $\mathrm{HV}$ and $\mathrm{HH}$ gates produce identical results.  The dashed line shows $\mathcal{F} = 0.9999$ for reference.}
    \label{fig:XXSims}
\end{figure}

Inter-atomic two-qubit gates can be constructed from the standard tools of trapped ion processing as long as they utilize the \emph{qubit} Paulis ($\hat{\sigma}^{(i)}_\mrm{H}$ and $\hat{\sigma}^{(i)}_\mrm{V}$) implemented simultaneously, as described above \cite{SI}.
For example, an \textsc{xx} gate between two monoencoded ions is typically implemented via the M\o{}lmer-S\o{}rensen (MS) interaction \cite{Sorensen1999Quantum}, wherein a bichromatic drive couples the qubits using a normal mode of motion as a Bosonic quantum bus.  
For polyqubit encodings, this gate is implemented by applying $2^{p-1}$ copies of the MS interaction that couple each atomic state comprising the participant qubit $\ket{0}$ to a unique atomic state comprising the participant qubit state $\ket{1}$.
Because these MS interactions must couple to the same Bosonic mode, they do not commute and must be applied simultaneously, with matched strength and phase, to ensure that the spectator qubits are undisturbed.

As an example, an inter-atomic \textsc{xx} gate between participant qubits, $d_k \in \{H,V\}$, hosted in two ions (indexed by $k$) with a $p = 2$ polyqubit encoding is realized by simultaneously applying a matched MS interaction with effective resonant sideband Rabi frequencies $\atomicRabi{m}{n}$ to multiple pairs of atomic states in each ion \cite{SI}. 
The calculated time evolution of the qubit state populations during this interaction are shown in Fig.~\ref{fig:XXSims}a, as points, for the spectator qubits in arbitrary initial states and the participant qubits initialized in $\ket{0}_{d_1}\otimes\ket{0}_{d_2}$. 
The solid lines show the evolution 
of the traditional MS interaction on monoencoded ions under the same conditions.

As can be seen from the agreement between the dashed lines and the points, the polyqubit entangling gate on the participant qubits does not interfere with information stored in the spectator qubits as long as the pairwise interactions within each atom are identical \cite{SI}.
Fig.~\ref{fig:XXSims}b shows the effect in a mismatch in Rabi frequency for the \textsc{xx} gate on participant qubits in two polyencoded atoms.
In this figure, the points, connected by lines to guide the eye, give the average fidelity for creating a Bell state with the interaction from the initial state $\ket{0}_{d_1}\otimes\ket{0}_{d_2}$, averaged over 100 random initial states of the spectator qubits, while the error bars show the standard deviation. 
By matching the Rabi frequencies to roughly 0.5\%, fidelities greater than $\mathcal{F}=0.9999$ are possible. 

\begin{figure}
    \centering
    \includegraphics[width = 0.38\textwidth]{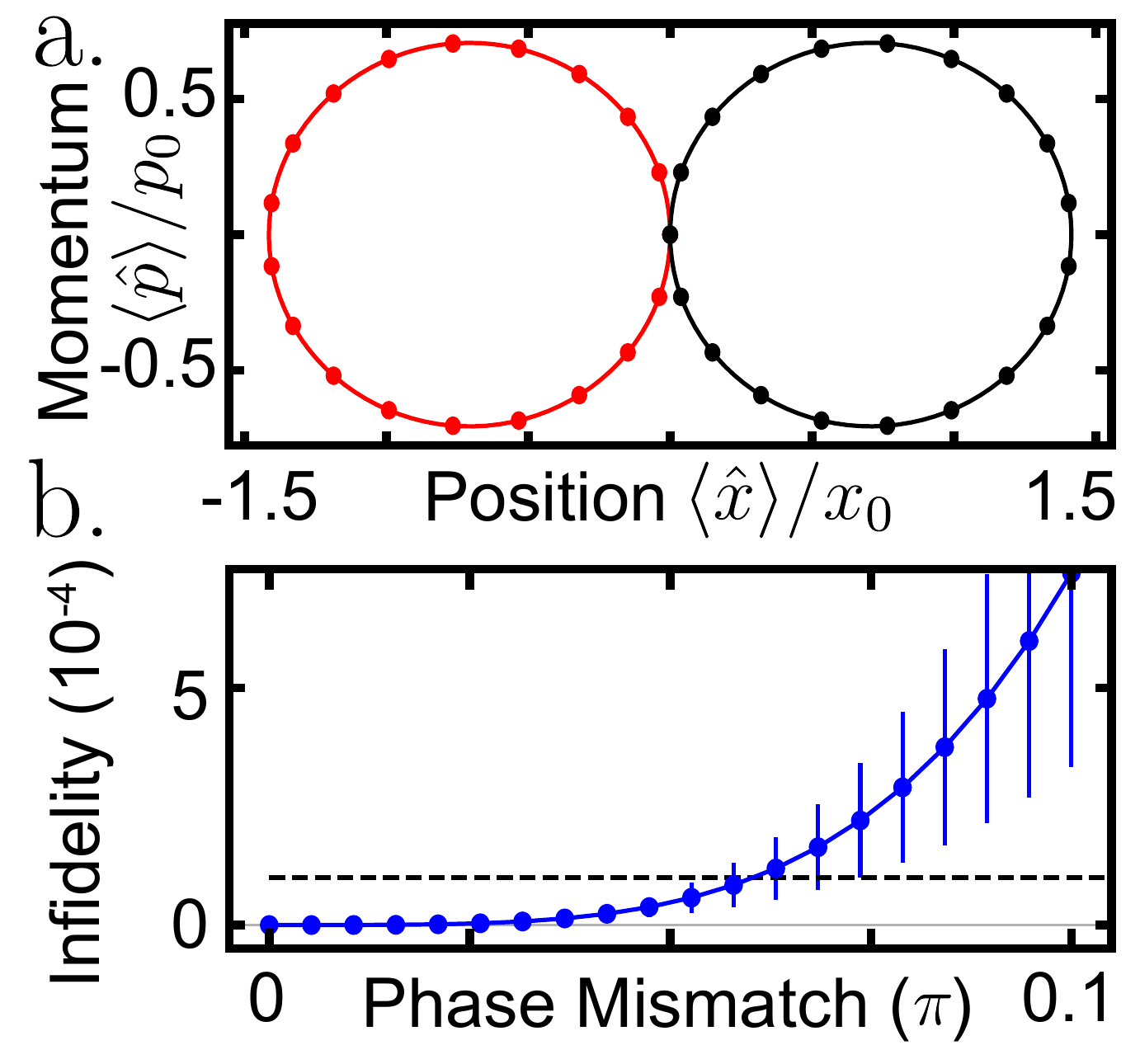}
    \caption{(a) Calculated normal mode phase-space trajectories for two polyqubits during an inter-atomic \ZZ\ gate. (b) Mismatched beatnote phases on the interactions applied within each individual atom can degrade the fidelity, but only for large mismatches.}
    \label{fig:ZZSims}
\end{figure}

In similar fashion, \textsc{zz} gates \cite{Leibfried2003Experimental,Roos2008} can be constructed for polyqubits by implementing the traditional monoqubit scheme with polyencoded qubit operators \cite{SI}.  
This can be accomplished, for example, by applying a highly-state-selective AC Stark shift that is modulated near, but intentionally detuned from  a normal mode frequency \cite{Baldwin2021HighFidelity}, similar to Fig.~\ref{fig:SPAM}b and c.  
Figure \ref{fig:ZZSims}a shows the phase-space trajectory of two polyqubits during a \textsc{zz} gate on target qubits for an arbitrary initial state of the spectator qubits.
The evolution within the participant qubit subspaces is identical to that expected for monoencoded qubits.
The points in Fig.~\ref{fig:ZZSims}b show the fidelity for creating a Bell state from the initial state $\ket{X}_{d_1}\otimes\ket{X}_{d_2}$ of the participant qubits, averaged over 100 random initial states of the spectator qubits, as a function of mismatch in beatnote phase between the two atomic interactions within each atom, while the error bars show the standard deviation.  The fidelity is relatively insensitive to beatnote phase mismatch within each atom.

While we have focused on trapped ion processors as a concrete and accessible example for polyqubit processing, the same idea is generally applicable to other systems.  
For example, neutral atom processors with Rydberg gates can potentially benefit from polyencoding.  
In this case, the SPAM would be furnished by a Rydberg gate with an ancilla atom that uses simultaneous driving of multiple transitions within the polyencoded atom to ensure that only the participant qubit is sensitive to the gate. 
As long as individual qubits within a polyencoded atom can be independently manipulated, polyqubit processing can be built from modified versions of the schemes used for monoqubit processing.

The polyqubit encoding presented here is in principle extendable to large values of $p$ if the host system has enough stable states.  
System such as molecules may be ideal in this respect, as they can have many stable, nonmagnetic states that can be controlled using electric dipole transitions in the microwave spectrum. 
Further, the vibrational degrees of freedom of molecules may provide convenient Hilbert subspaces for realizing polyencodings as they can be approximately factorable as separate degrees of freedom.
For atomic systems, it appears feasible to encode with $p \le 3$ by using species with nuclear spin. 
Even for these low values of $p$, however, polyqubit operation could provide ancillary benefits beyond increasing qubit number.  
These include the replacement of some inter-atomic gates by intra-atomic gates (which are likely to have significantly higher fidelity), reduced shuttling needs, and potentially increased connectivity in range-limited processors. 

While polyqubit processing has no more computational power than processing using equally-dimensioned qudits, it may have some practical advantage as it  requires controls that are essentially already present in trapped ion quantum processors.
Further, factoring a qudit Hilbert space into polyqubits provides a qudit-dimension independent means to process with current algorithms designed for qubits, including QEC.
As such, the resource cost associated with converting a binary machine to polyqubit mode to gain this multiplicative increase in the qubit number may prove to be a good bargain in the QHL era.

\subsubsection*{Acknowledgements}
This work was supported by the ARO under Grants No.\ W911NF-19-1-0297 and No.\ W911NF-20-1-0037, the AFOSR under Grant No.\ FA9550-20-1-0323, and the NSF under Grants No.\ PHY-2110421, No.\ PHY-2207985, and No.\ OMA-2016245.  WCC acknowledges support from the Gordon and Betty Moore Foundation's EPI initiative.
ERH acknowledge support from the NSF CCI Phase I grant 2221453.
We thank Paul Hamilton and Isaac Chuang for discussions.

\bibliography{MultiqubitEncoding}
\bibliographystyle{unsrt}

\clearpage

\onecolumngrid

\section{Supplemental Materials}
\section{Inter-atomic two-qubit gates}
\renewcommand{\theequation}{S.\arabic{equation}}
\setcounter{equation}{0}
Inter-atomic two-qubit gates can be achieved by essentially employing two copies of the interaction needed to effect two-qubit gates in monoqubit encodings. 
In this way, two-qubit gates between two $\mrm{H}$ or two $\mrm{V}$ qubits in different atoms, as well as between an $\mrm{H}$ qubit in one atom with a $\mrm{V}$ qubit in another, can be realized. 
As the two commonly-employed two-qubit interactions for trapped ion quantum process are $\hat{\sigma}^{(X,1)} \hat{\sigma}^{(X,2)}$ and $\hat{\sigma}^{(Z,1)} \hat{\sigma}^{(Z,2)}$, we discuss each in turn. 

\subsection{Inter-atomic \XX\ gates}
Similar to the single qubit gates, the two-qubit \XX\ gate between distinct polyqubit-encoded atoms is realized by implementing two \XX\ gates on the supporting atomic states.
Crucially, unlike the single qubit case, these two interactions driven on each atom must be performed simultaneously, using the same mode of motion, to ensure this is not accompanied by a collateral two-qubit operation between the participant and spectator qubits in the single atoms.

As an example, a $\hat{\sigma}_\mrm{V}^{(X)}\hat{\sigma}_\mathrm{V}^{(X)}$ operation can be realized using M\o{}lmer-S\o{}rensen interactions \cite{Sorensen1999Quantum} where motion-sensitive transitions between atomic eigenstates are driven near resonant with the first red and blue motional sidebands.
Thus, to realize the $\hat{\sigma}_\mrm{V}^{(X)}\hat{\sigma}_\mrm{V}^{(X)}$ interaction two such transitions are driven between  $\{\Aket{0},\Aket{1}\}$ and  $\{\Aket{2},\Aket{3}\}$.
In the interaction picture with respect to the atomic and the harmonic oscillator Hamiltonians and making the Lamb-Dicke approximation and the rotating wave approximation (RWA), the effective Hamiltonian~\cite{James2007effective} describing the evolution due to the laser fields is:
\begin{align}
    H_\mrm{VV}^{(XX)} &= \sum_{\alpha=1}^2 \eta_{\alpha}\left( \frac{\atomicRabi{0}{1}_\alpha}{2} \,\atomicPauli{+,\alpha}{0}{1}  (\hat{a}e^{\imath \delta t} + \hat{a}^\dagger e^{-\imath \delta t}) e^{\imath \Delta\phi_{\mathtt{\underline{0}\hspace{0.1mm}\underline{1}}}} + \frac{\atomicRabi{2}{3}_\alpha}{2}\, \atomicPauli{+,\alpha}{2}{3} (\hat{a}e^{\imath \delta t} + \hat{a}^\dagger e^{-\imath \delta t}) e^{\imath\Delta\phi_{\mathtt{\underline{2}\hspace{0.1mm}\underline{3}}}} +  \textrm{H.c.}\right),
    \label{eqn:XXV_td}
\end{align}
where $\delta$ is the detuning from the motional sidebands including the laser-induced Stark shifts and if the gate is being driven by stimulated Raman transisions, the $\atomicRabi{m}{n}$ contain the resonant, single-photon Rabi frequencies of the individual beams and their detunings from atomic resonance.
In order to isolate the participant qubit, we will from this point  assume that the two interactions with the lasers have the same strength, $\eta_{\alpha}\atomicRabi{0}{1}_\alpha = \eta_{\alpha}\atomicRabi{2}{3}_\alpha \equiv g_\alpha$, and phase, $\Delta\phi_{\mathtt{\underline{0}\hspace{0.1mm}\underline{1}}} = \Delta\phi_{\mathtt{\underline{2}\hspace{0.1mm}\underline{3}}} = 0$, which allows us to write this in terms of qubit Paulis using Eq.~(\ref{eq:AtomicPaulis}),
\begin{align}
    H_\mrm{VV}^{(XX)} &= \sum_{\alpha=1}^2  \frac{g_\alpha}{2} \,\hat{\sigma}^{(X,\alpha)}_\mrm{V}  (\hat{a}e^{\imath \delta t} + \hat{a}^\dagger e^{-\imath \delta t}). \label{eq:HVVXX}
\end{align}

In the limit where the phonon excitations can be adiabatically eliminated, this Hamiltonian leads to evolution described by the effective Hamiltonian \cite{James2007effective}:
\begin{align}
     H_\mrm{VV,eff}^{(XX)} &\approx \frac{g_1 g_2}{2\delta}\left(
\atomicPauli{X,1}{0}{1}\atomicPauli{X,2}{0}{1} + \atomicPauli{X,1}{0}{1}\atomicPauli{X,2}{2}{3}  + \atomicPauli{X,1}{2}{3}\atomicPauli{X,2}{0}{1}  + \atomicPauli{X,1}{2}{3}\atomicPauli{X,2}{2}{3} \right) = \frac{g_1 g_2}{2\delta}\sigma_\mrm{V}^{(X,1)}\sigma_\mrm{V}^{(X,2)} \label{eqn:XXVa}
\end{align}
The form of this result can be readily understood.
The first and fourth terms in Eqn.~\ref{eqn:XXVa} arise in the usual way for a M\o{}lmer-S\o{}rensen gate, while the middle terms arise from the excitation of one ion by one stimulated Raman path and the excitation of the other ion by the other stimulated Raman path. 
These four processes together ensure no entanglement between the $\mrm{H}$ and $\mrm{V}$ qubits and yield the desired interaction.

In much the same fashion as MS gates for monoencoded atoms, this effective Hamiltonian (\ref{eqn:XXVa}) serves to highlight a limiting case that can be used to develop intuition, but a realistic gate is likely to operate in a regime where the adiabatic elimination of phonons is not valid.  In this case, the gate time must be chosen to decouple the qubit from the motional evolution, potentially in multiple modes simultaneously.  Once the interactions within each polyencoded atom are matched to one another, Equation (\ref{eq:HVVXX}) is identical to the form used for monoencoded atoms, which gives polyqubit processors access to the tools that have been developed for monoqubit systems to manage multiple modes and operate gates in a robust fashion.

Figure~\ref{fig:XXSims} shows a numerical solution of the time-dependent Schr\"odinger equation under Eq.~(\ref{eqn:XXV_td}).
Dashed lines are for two ions starting in either $\Aket{00}$ or $\Aket{22}$, showing that the evolution is the same as an MS gate for monoencoded qubits.
The dots are the same evolution but with an initial state of $\Hket{\alpha\beta} \otimes \Vket{00}$, where for each point a different initial state is chosen for the $\mrm{H}$ qubits in ions 1 or 2. 
This illustrates that there is no residual entanglement involving the spectator qubits as the interaction acts only on the participant qubits. 

Similarly, two-qubit gates on $\mrm{H}$ qubits proceed by driving near resonant with the first red and blue motional sidebands of the $\{\Aket{0},\Aket{2}\}$ and  $\{\Aket{1},\Aket{3}\}$ transitions.
The resulting effective Hamiltonian is 
\begin{align}
    H_\mrm{HH,eff}^{(XX)} \approx \frac{g_1g_2}{2\delta}\sigma_\mrm{H}^{(X,1)}\sigma_\mrm{H}^{(X,2)}.
\end{align}

Finally, assuming single ion addressing is available, inter-atomic two-qubit gates between dissimilar qubits (such as $\mrm{HV}$) are possible via illuminating each ion with the appropriate radiation.

\subsection{Inter-atomic \ZZ\ gates}
For monoqubit encoded ions, the differential Stark shift of the two atomic states comprising a qubit provides an optical force on a trapped ion that depends on the qubit state.
When used with a bichromatic laser field that has a beat frequency ($\omega_\mrm{b}$) near a motional mode freqeuncy of the trapped ions ($\omega$), the trapped ions can experience an effective  $\hat{\sigma}^{(Z,1)}\hat{\sigma}^{(Z,2)}$ interaction~\cite{Roos2008}. 
Similarly, $\hat{\sigma}^{(Z,1)}\hat{\sigma}^{(Z,2)}$ polyqubit gates can be realized by applying two bichromatic laser fields that produce a differential optical force within the appropriate \emph{pairs} of states. 
As an example, $\hat{\sigma}_\mrm{V}^{(Z,1)}\hat{\sigma}^{(Z,2)}_\mrm{V}$ interaction is realized by applying a bichromatic field that produces a differential optical force between states $\{\Aket{0},\Aket{1}\}$ and another bichromatic field that produces a differential optical force between states $\{\Aket{2},\Aket{3}\}$, detuned from the motional mode freqeuncy by $\delta \equiv \omega_\mathrm{B} - \omega$.
In the interaction picture with respect to the harmonic oscillator and the atomic Hamiltonians and after the RWA and Lamb-Dicke approximations, the evolution is described by the effective Hamiltonian:
\begin{align}
   H_\mrm{VV}^{(ZZ)} & =  \sum_{\alpha = 1}^2 \eta_\alpha \bigg( \frac{\Delta\atomicRabi{0}{1}_\alpha}{2} 
   \atomicPauli{Z,\alpha}{0}{1}
     \left(\hat{a} e^{\imath(\delta t + \phi_{\mathtt{\underline{0}\hspace{0.1mm}\underline{1}}})} + \hat{a}^\dagger e^{-\imath(\delta t + \phi_{\mathtt{\underline{0}\hspace{0.1mm}\underline{1}}})} \right) + \frac{\Delta\atomicRabi{2}{3}_\alpha}{2}
    \atomicPauli{Z,\alpha}{2}{3} \left(\hat{a} e^{\imath(\delta t + \phi_{\mathtt{\underline{2}\hspace{0.1mm}\underline{3}}})}   + \hat{a}^\dagger e^{-\imath(\delta t + \phi_{\mathtt{\underline{2}\hspace{0.1mm}\underline{3}}})}\right) \bigg).\label{eq:HZZVVSI}
\end{align}
where $\Delta \atomicRabi{m}{n}_\alpha$ is the differential AC Stark shift amplitude between atomic states $\Aket{m}_\alpha$ and $\Aket{n}_\alpha$ and $\phi_{\mathtt{\underline{m}\hspace{0.1mm}\underline{n}}}$ is the phase of the beatnote on the differential AC Stark shift of states $\Aket{m}_\alpha$ and $\Aket{n}_\alpha$.  As with the \XX\ gate, the two strengths within each ion should be matched, and we will assume that $\eta_\alpha \Delta\atomicRabi{0}{1}_\alpha = \eta_\alpha \Delta\atomicRabi{2}{3}_\alpha \equiv g_\alpha$ and that the phase of the low-frequency beatnotes are also equal, $\phi_{\mathtt{\underline{0}\hspace{0.1mm}\underline{1}}} = \phi_{\mathtt{\underline{2}\hspace{0.1mm}\underline{3}}} = 0$.  In order to represent the full Hamiltonian by equation (\ref{eq:HZZVVSI}), we also require that the common-mode (as opposed to differential) AC Stark shifts of the $\{\Aket{0},\Aket{1}\}$ states is the same as the $\{\Aket{2},\Aket{3}\}$ states, though a known mismatch could be correct by a $\hat{\sigma}_H^{(Z)}$ rotation. This allows us to apply Eq.~(\ref{eq:AtomicPaulis}) to identify this as
\begin{align}
     H_\mrm{VV}^{(ZZ)} & =  \sum_{\alpha = 1}^2 \frac{g_\alpha}{2} \hat{\sigma}_\mrm{V}^{(Z,\alpha)}
     \left(\hat{a} e^{\imath \delta t} + \hat{a}^\dagger e^{-\imath \delta t}  \right)
\end{align}

Once again taking the limit in which the detuning $\delta$ is large compared to the interaction strength so that phononic excitations can be adiabatically eliminated, This Hamiltonian leads to evolution described by the effective Hamiltonian:
\begin{align}
    H_\mrm{VV,eff}^{(ZZ)} & \approx \frac{g_1g_2}{2\delta}
    \left(
    \atomicPauli{Z,1}{0}{1}\atomicPauli{Z,2}{0}{1} + \atomicPauli{Z,1}{0}{1}\atomicPauli{Z,2}{2}{3}  + \atomicPauli{Z,1}{2}{3}\atomicPauli{Z,2}{0}{1}  + \atomicPauli{Z,1}{2}{3}\atomicPauli{Z,2}{2}{3} \right)
    = \frac{g_1g_2}{2\delta}\sigma_V^{(z,1)}\sigma_V^{(z,2)} \label{eqn:ZZVa}
\end{align}
As with the $\sigma^{(X)}\sigma^{(X)}$ interaction, the form of this result can be readily understood.
The first and fourth terms in Eq.~(\ref{eqn:ZZVa}) arise in the same manner as a monoencoded $\sigma^{(Z)}\sigma^{(Z)}$ gate, while the middle terms arise from the cross terms. 
These four processes together ensure no entanglement between the $\mrm{H}$ and $\mrm{V}$ qubits and yield the desired interaction.

\section{Intra-atomic Deutsch and Toffoli gates for $p=3$}

When $p\ge 1$, interactions that involve only a single atomic Pauli operator can be nontrivial in the qubit basis.
For example, consider the case of $p=3$ that is depicted schematically in Figure~\ref{fig:StatesBallAndStick}d.  If a $\Theta$ pulse is driven using $\atomicPauli{X}{6}{7}$, the evolution operator can be written
\begin{align}
    \exp \left(-i \frac{\Theta}{2}\atomicPauli{X}{6}{7}\right) = \left( \begin{array}{cc} \idty_6 & 0 \\ 0 & \exp \left(-i \frac{\Theta}{2}\sigma^{(X)} \right)  \end{array} \right),
\end{align}
where $\idty_6$ is the $6\times 6$ identity operator.  This performs a rotation on the $\ket{110} \leftrightarrow\ket{111}$ subspace.  Next, we can shift the relative phase between those two states and the rest of the space by, for instance, applying other atomic Pauli interactions (either simultaneously or sequentially)
\begin{align}
    \hat{U}_{\mathtt{\underline{6}\hspace{0.1mm}\underline{7}}} \equiv e^{i \frac{\pi}{8}} \exp \left( -i \frac{\pi}{8} \left(2 \atomicPauli{Z}{5}{7} + 2 \atomicPauli{Z}{4}{6} + \atomicPauli{Z}{1}{5} + \atomicPauli{Z}{3}{7} + \atomicPauli{Z}{0}{4} + \atomicPauli{Z}{2}{6} \right)\right),
\end{align}
and the resulting rotation is a Deutsch gate \cite{Deutsch1989}:
\begin{align}
    \hat{U}_{\mathtt{\underline{6}\hspace{0.1mm}\underline{7}}} \exp \left(-i \theta\atomicPauli{X}{6}{7}\right) = \left( \begin{array}{cccccccc} 1 & 0 & 0 & 0 & 0 & 0 & 0 & 0 \\ 0 & 1 & 0 & 0 & 0 & 0 & 0 & 0 \\ 0 & 0 & 1 & 0 & 0 & 0 & 0 & 0 \\ 0 & 0 & 0 & 1 & 0 & 0 & 0 & 0 \\ 0 & 0 & 0 & 0 & 1 & 0 & 0 & 0\\ 0 & 0 & 0 & 0 & 0 & 1 & 0 & 0 \\ 0 & 0 & 0 & 0 & 0 & 0 & i \cos\left( \theta \right) & \sin\left( \theta \right) \\ \hspace{5mm}0\hspace{5mm} & \hspace{5mm}0\hspace{5mm} & \hspace{5mm}0\hspace{5mm} & \hspace{5mm}0\hspace{5mm} & \hspace{5mm}0\hspace{5mm} & \hspace{5mm}0\hspace{5mm} & \sin\left( \theta \right) & i\cos\left( \theta \right)  \end{array} \right) = \mathbbm{D}_3(\theta).
\end{align}
A 3-qubit Toffoli gate is the special case $\mathbbm{D}_3(\pi/2)$.  Extensions of these ideas to perform other types of gates (\textit{e.g.} \textsc{swap}, \textsc{cswap}, \textsc{cz}, etc.) can be obtained by straightforward substitutions.  For example, an intra-atomic $p$-qubit Deutsch gate (of which \textsc{cnot} and $p$-qubit Toffoli gates are special cases) can be performed by driving a rotation on $\ket{11\cdots 10} \leftrightarrow \ket{11\cdots 11}$ followed by phase shifts,
\begin{align}
    \mathbbm{D}_p(\theta) =  e^{i \frac{\pi}{2^p}} \exp \left(-i \frac{\pi}{2^p} \sum_\ell^{2^p - 3} \hat{s}^{(Z)}_{\underline{\ell_{}^{}},\underline{f(\ell)}} \,  \right) \exp \left(-i \theta \hat{s}^{(X)}_{\underline{2^p - 2},\underline{2^p - 1}} \right) 
\end{align}
where $f(\ell) \equiv 2^p - 2 + (\ell \mod 2)$.

\end{document}